\documentclass[runningheads]{llncs}
\usepackage{graphicx}
\usepackage{todonotes}
\usepackage{algpseudocode}
\usepackage[Algorithm]{algorithm}
\usepackage{mathtools}
\usepackage{amsmath}
\usepackage{amssymb}
\usepackage{wasysym}
\usepackage{cite}
\usepackage{bibnames}
\usepackage{multirow}
\usepackage{subcaption}
\usepackage{color, colortbl}
\usepackage{sidecap}
\usepackage{enumitem}

\newcommand{\ignore}[1]{}
\newcommand{\use}{\texttt{USE}}
\newcommand{\iacd}{\texttt{IACD}}
\newcommand{\university}{\texttt{Universities}}
\newcommand{\proprietary}{\texttt{Proprietary}}
\newcommand{\airport}{\texttt{Airport}}
\newcommand{\bleu}{\texttt{BLEU}}
\newcommand{\cosine}{\texttt{Cosine}}
\newcommand{\meteor}{\texttt{METEOR}}

\newcommand{\slicing}{\texttt{Slicing}}
\newcommand{\random}{\texttt{Random}}

\newcommand{\iacdt}{IACD}
\newcommand{\universityt}{Universities}
\newcommand{\proprietaryt}{Proprietary}
\newcommand{\airportt}{Airport}
\newcommand{\bleut}{BLEU}
\newcommand{\cosinet}{Cosine}
\newcommand{\meteort}{METEOR}
\newcommand{\precisiont}{precision@3}
\newcommand{\recallt}{recall@3}
\newcommand{\slicingt}{Slicing}
\newcommand{\randomt}{Random}

\newcommand{\slice}{$5$}
\newcommand{\topk}{$3$}

\newcommand\real{\mathbb{R}}

\begin{document}
%

\title{Augmenting Modelers with Semantic Autocompletion of Processes}

\author{Maayan Goldstein \and Cecilia Gonz{\'{a}}lez-\'{A}lvarez}

\institute{Nokia Bell Labs\\
\email{\{maayan.goldstein, cecilia.gonzalez\_alvarez\}@nokia-bell-labs.com}
}

\maketitle              
\begin{abstract}
Business process modelers need to have expertise and knowledge of the domain that may not always be available to them. 
Therefore, they may benefit from tools that mine collections of existing processes and recommend element(s) to be added to a new process that they are constructing. 
In this paper, we present a method for process autocompletion at design time, that is based on the semantic similarity of sub-processes.
By converting sub-processes to textual paragraphs and encoding them as numerical vectors, we can find semantically similar ones, and thereafter recommend the next element.
To achieve this, we leverage a state-of-the-art technique for embedding natural language as vectors.
We evaluate our approach on open source and proprietary datasets and show that our technique is accurate for processes in various domains.

\end{abstract}
%
%
%


\section{Introduction}
\label{sec:intro}

\ignore{
\begin{itemize}
    \item Business processes are used across different domains and organizations
    \item Reuse and automation are paramount to enable faster development of these models and catch bugs
    \item One way to enable automation is by prediction of what should come next
    \item This is already used successfully in software, and initial attempts have already been done for business processes~\cite{wang2019process}
    \item We suggest a new technique that is using Universal Sentence Encoder~\cite{cer2018universal} 
    \item Why Universal Sentence Encoder? It has been successfully applied in...(TBC)
\end{itemize}

Using past experience to suggest next steps in a workflow design eliminates much of the guesswork and time-consuming effort of reading component API documents trying to determine what options are available
}

Business processes are used across different domains and organizations. Commercial companies and the public sector alike have adopted 
process models as a means for visualizing and executing their business logic.
As the volume of existing process models in an organization increases, it becomes evident that reuse and automation are paramount to enable faster design of high-quality models~\cite{fellmann2018business,fellmann2015requirements}.

\begin{figure}[t]
  \begin{center}
    \includegraphics[width =
       0.9\columnwidth] {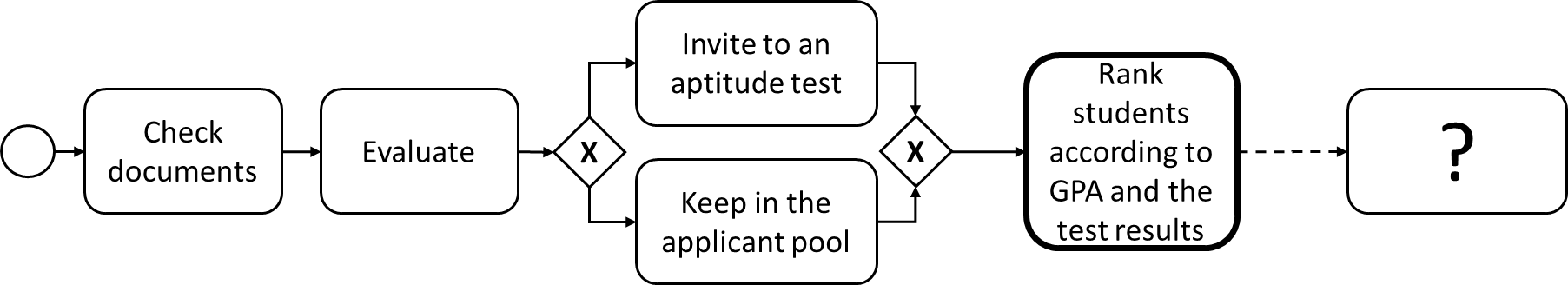}  
  \end{center}
  \caption{Autocompletion problem example: the modeler is interested in having a recommendation for elements that could follow the task ``Rank students according to GPA and the test results'', in a process excerpt  from~\cite{antunes2015process}.  
  }
  \label{fig:problem}
  \vspace{-0.2cm}
\end{figure}

One way to enable automation is by suggesting what should come next in a model that is being constructed, as exemplified in Figure~\ref{fig:problem}. Using knowledge from previous experience to recommend the next steps in a process design saves the modeler  much of the guesswork and time-consuming effort of reading documentation trying to determine what options are available.

Autocompletion systems have become popular in recent years as they boost productivity and improve quality.  
For instance, modern e-mail applications suggest how to automatically complete sentences for the user, saving time, improving grammar and style, and avoiding typos~\cite{cornea2014providing}. 
Software developers that are using tools for automatic completion of their code~\cite{tab9}, benefit from richer development experience leading to fewer bugs and better code reuse.
Our goal is to provide process modelers with a similar user experience.

There are two main driving forces for the accelerated development of autocompletion tools in the recent years:
first, the growth in the amount of text and open source code available on the web, and second, recent advances in deep learning that benefit from large volumes of information.
These two factors contribute to the widespread adoption of deep learning models that are highly accurate and easily transferable to various domains~\cite{floridi2020gpt,lavie2007meteor}.

However, autocompletion of business process models has not experienced a breakthrough yet, as vast repositories of open-source models do not exist. 
As machine learning techniques that train recommendation systems from scratch require large amounts of data, they are  inapplicable for process autocompletion.


Several past attempts to solve the autocompletion problem focused on syntactic information, such as the structure of process model graphs, and did not take into account the semantic meaning of the processes and their fragments~\cite{fellmann2018business, li14, hornung2007rule,born2008auto,zhang09}. 
Other researchers investigated semantic-based approaches~\cite{wieloch2011autocompletion,wang2019process,koschmider2011recommendation,koschmider2010social}, however, semantic similarity between sub-processes determined with modern deep learning techniques has not been fully explored yet.

\vspace{-0.4cm}
\subsubsection{Our contributions.} In this paper, we propose to use semantic similarities among processes to enable the autocompletion of the next element(s) at design time. 
Our approach takes into account the limited availability of data in the field by leveraging pre-trained models for natural language processing (NLP). It also overcomes the obstacle of handling elements that bear similar meaning, but somewhat different textual description, by matching tasks with similar labels rather than exact matches only.
Our solution transforms sequences of process elements into paragraphs of text and represents them as sentence embeddings, which are learned representations of text that capture semantic information as vectors of real numbers.

The computed embeddings of element sequences can be compared to each other via techniques that measure the distance between vectors. 
Thus, given a partially completed process, we can find processes in a repository of existing business processes that are semantically similar to it. 
Thereafter, we can recommend the most likely element to be added to the process, based on these similar processes. 

We also present a framework for evaluation of next-element recommendation systems, filling the gap of previous works on autocompletion.
We evaluate the effectiveness of our approach using
metrics widely utilized in NLP and recommendation systems.

 \section{Semantic Autocompletion in a Nutshell}


Semantic autocompletion aims at recommending process elements based on semantic similarity of the process being developed to other processes from a given repository.

\begin{figure}[h]
  \begin{center}
    \includegraphics[width =
        \columnwidth] {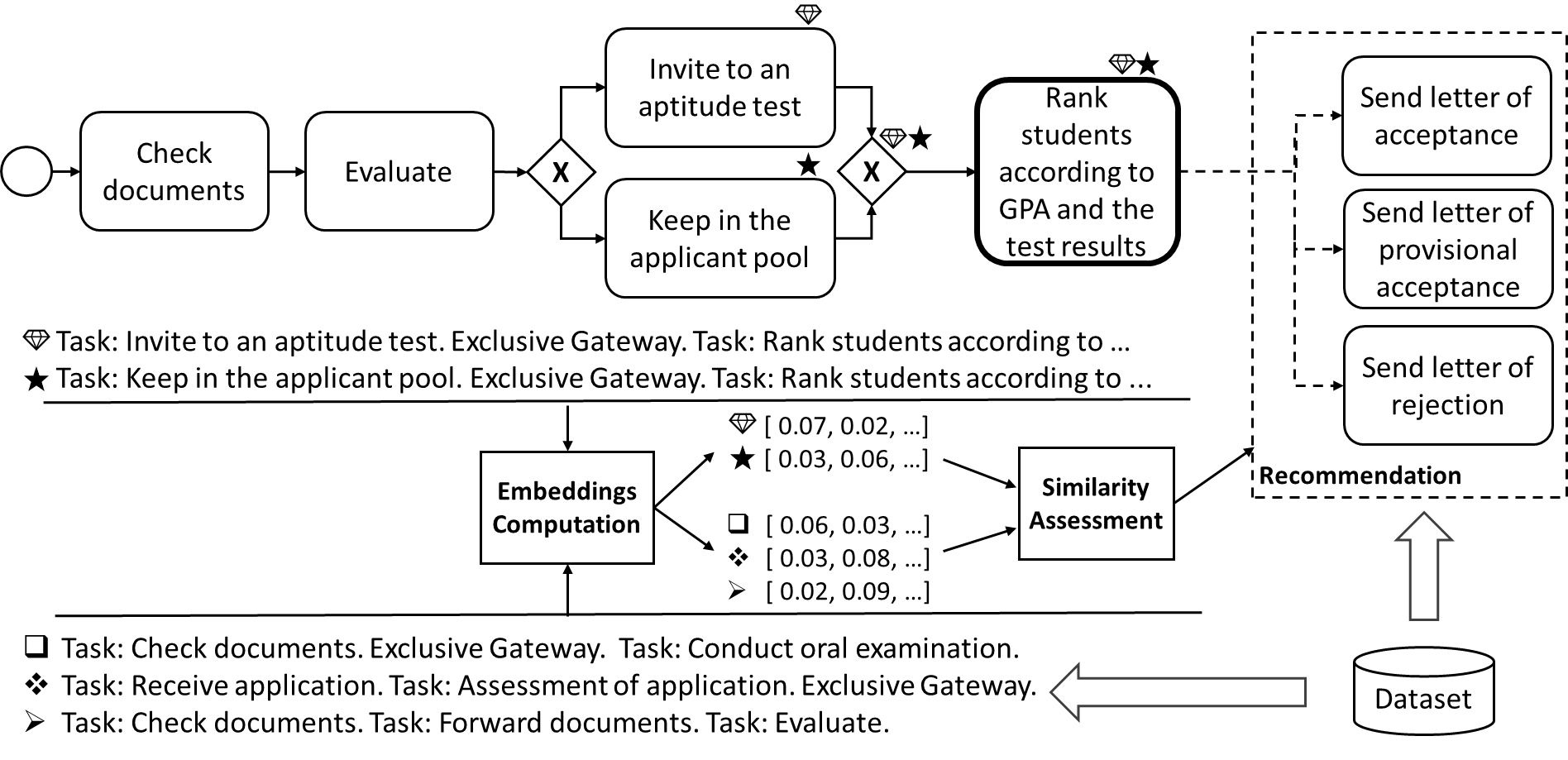}  
  \end{center}
  \caption{Overview of our approach for the autocompletion of a partial process model. 
  Based on the most semantically similar paths in the input dataset, up to \topk{} top recommendations are shown to the modeler. 
  }
  \label{fig:process}
  \vspace{-0.2cm}
\end{figure}

Our autocompletion engine works following the procedure illustrated in Figure~\ref{fig:process}. 
The unfinished process at the top is an excerpt from a process taken from the university admissions dataset~\cite{antunes2015process} and represents the application procedure for master students in Frankfurt university. 

 The autocompletion engine may suggest the modeler that is developing this process which is the most likely next element to the last one added. 
In our example, the last element is the task ``Rank students according to GPA and the test results'', marked with a bold outline.  
To autocomplete the process, our algorithm first traverses all the sub-processes of a predefined length leading to that specific task.
In the example, we consider sub-processes of length three, resulting in the two paths marked with a star and a diamond.
Next, we convert each sub-process into a paragraph of text by concatenating the labels and type names of the elements in the sub-process. 

Third, for each paragraph, we compute its vector embedding, such that an arbitrary length text is converted to a fixed length numerical vector. 
Then, each computed embedding of the target process is compared to the embeddings of all the sub-processes from an existing dataset of processes (exemplified at the bottom of the figure), via a similarity metric. 
Finally, if there are sub-processes in the input dataset that are semantically similar to the sub-process of interest, the top matching recommendations are shown to the user. 
These recommendations are the elements that appeared in the dataset for other universities right after the most similar paths (that is, were connected to those paths via a connector).

\section{Preliminaries}
\label{sec:preliminaries}

Finding similarities between processes is key to our autocompletion strategy, that is fully explained in Section~\ref{sec:approach}.
This section serves as a background on how we apply vector representations of text to our problem. 
We also provide formal definitions for concepts used throughout the rest of the paper.

\subsection{Universal Sentence Encoder, Embeddings and Similarity}

The way we convert paragraphs of text into vectors relies on
a pre-trained deep learning model called the Universal Sentence Encoder (\use{})~\cite{cer2018universal}.
\use{} encodes text as high-dimensional numerical vectors that can be used for text classification, semantic similarity assessment, clustering, and other tasks that involve natural language processing (NLP).
Intuitively, embeddings are a mathematical representation of the semantics of the sentences. 

The main advantage of embeddings is that sentences of an arbitrary length are transformed into vectors of real numbers of the same length.
This enables comparison of pairs of sentences by means of computing a similarity score between vectors representing the sentences.

Let $U$ be the \use{} model~\cite{cer2018universal}.
Given an input sentence $p$ which is a list of words in English, we define \textit{sentence embedding} as follows:

\ignore{
\begin{definition}[Sentence embedding]
	\label{def:sentence_embedding}
	For an input sentence $s$, the embedding of $s$ is given by 
	$\vec{e(s)} = \hat{U}(s)$. 
\end{definition}
Where $\hat{U}(s) \in \real^m$ is the \textit{valuation} of $U$ on $s$, and $m$ is the length of the embedding vector.  We will use the notation \textbf{e} instead of 	$\vec{e(s)}$ whenever $s$ is clear from the context.
}

\begin{definition}[Sentence and paragraph embedding]
	\label{def:sentence_embedding}
	For an input sentence (or paragraph) $p$, the embedding of $p$ is given by 
	$ \textbf{p} = U(p)$. 
\end{definition}
Where $U(p) \in \real^n$, and $n$ is the length of the embedding vector. Note that a paragraph that contains multiple sentences will be also encoded as a vector of length $n$.

\ignore{
\begin{definition}[Sentence and paragraph embedding]
	\label{def:sentence_embedding}
	For an input sentence (or paragraph) $p$, the embedding of $p$ is given by 
	$ \textbf{p} = \hat{U}(p)$. 
\end{definition}
Whereas $\hat{U}(p) \in \real^m$ is the \textit{valuation} of $U$ on $p$, and $m$ is the length of the embedding vector. Note that a paragraph that contains multiple sentences will be encoded in the same manner as a vector of length $m$.
}



Once sentences are encoded as vectors, we can calculate how close those vectors are to each other, and use this information as a measure of how semantically similar the corresponding texts are. 
Cosine similarity is often used in NLP to compare embeddings, and is defined as follows: 


\begin{definition}[Cosine similarity]
	\label{def:cosine}
	Given embeddings  	$ \textbf{p}$ and	$ \textbf{q}$ for two sentences $p$ and $q$, the cosine similarity is computed as:
	\begin{equation} \label{eq:cosine}
\cos (\textbf{p},\textbf{q})= {\textbf{p} \cdot \textbf{q} \over \|\textbf{p}\| \|\textbf{q}\|} = \frac{ \sum_{i=1}^{n}{ p_i q_i} }{ \sqrt{\sum_{i=1}^{n}{p_i^2}} \sqrt{\sum_{i=1}^{n}{q_i^2}} }
\end{equation}
\end{definition}

\ignore{
\begin{definition}[Cosine similarity]
	\label{def:cosine}
	Given embeddings  	$ \vec{\textbf{s}}$ and	$ \vec{\textbf{q}}$ for two sentences $s$ and $q$, the cosine similarity is computed as:
	\begin{equation} \label{eq:cosine}
\cos (\vec{\textbf{s}},\vec{\textbf{q}})= {\vec{\textbf{s}} \cdot \vec{\textbf{q}} \over \|\vec{\textbf{s}}\| \|\vec{\textbf{q}}\|} = \frac{ \sum_{i=1}^{n}{{\bf s}_i{\bf q}_i} }{ \sqrt{\sum_{i=1}^{n}{({\bf s}_i)^2}} \sqrt{\sum_{i=1}^{n}{({\bf q}_i)^2}} }
\end{equation}
\end{definition}

\begin{definition}[Angular similarity]
	\label{def:angular}
	Given embeddings {\bf u} = $\vec{e(s_1)}$ and {\bf v} = $\vec{e(s_2)}$ for two sentences $s_1$ and $s_2$, the angular similarity is computed as:
\begin{equation}\label{eq:angular}
sim ({\bf u},{\bf v})= 1 - \arccos(\cos ({\bf u},{\bf v}))/ \pi
\end{equation}
\end{definition}
}

\ignore{
\begin{definition}[Angular similarity]
	\label{def:angular}
	Given embeddings $ \vec{\textbf{s}}$ and $\vec{\textbf{q}}$ for two sentences $s$ and $q$, the angular similarity is computed as:
\begin{equation}\label{eq:angular}
sim (\vec{\textbf{s}},\vec{\textbf{q}})= 1 - \arccos(\cos (\vec{\textbf{s}},\vec{\textbf{q}}))/ \pi
\end{equation}
\end{definition}
}

We now define a similarity matrix between two sets of sentence  embeddings $X = ({\bf x_1},...,{\bf x_r})$ and $Y = ({\bf y_1},...,{\bf y_m})$: 
\begin{definition}[Similarity matrix]
	\label{def:sim_matrix}
	$M(X, Y) = (cos({\bf x_i},{\bf y_j})), i=1,...,r; j=1,...,m$, where ${\bf x_1},...,{\bf x_r}$,  ${\bf y_1},...,{\bf y_m}$ are embedding vectors.
\end{definition}

\subsection{Process Model}

A process consists of a set of elements and connections between those elements, which can be described by a directed graph as follows:

\begin{definition}[Process as a directed graph]
	\label{def:process_graph}
	Let G=\{V, E, s, t\} where V is a set of elements, 
	E is the set of flows
	, $s \in V$ is the ``start'' event at which the process starts, and $t \in V$ is the ``end'' event at which the process terminates.
\end{definition}

Each node $v \in V$ can represent an activity, an event or a gateway used as a decision point. Each node may have a label and always has a type, that is $v = (label, type)$, whereas, optionally, $label = NULL$.

While making a recommendation for node $v$, we first need to extract all sub-processes of predefined length that end in $v$. We then compare these sub-processes to sub-processes extracted from the input dataset to find the most similar ones. We refer to these sub-processes as ``slices'':

\begin{definition}[Slice]
	\label{def:slice}
	Given a process $G=\{V, E, s, t\}$ and a number $n \in \mathbb{N}$, we say that $S_n=\{V_s, E_s\}$ is a slice of $G$ of length $n$, if $V_s \subseteq V$, $E_s \subseteq E$, $|V_s| = n$, and $S_n$ is a path graph.
\end{definition}

Note that since $S_n$ is a path graph~\cite{bondy1976}, its nodes can be topologically ordered such that we can later process the labels and the types of the nodes as if they were sentences following one another. We treat each slice as a paragraph of text, comprised of $n$ sentences. 

\subsection{Process Matching and Element Autocompletion}

Our solution makes its recommendations based on the best matching slice that it finds in its input dataset. That is, for the last node $v$ in an incomplete process, it extracts all the slices leading to that node, and for each such slice looks for the most similar slice in the input dataset. The best match is defined as follows:

\begin{definition}[Best match]
	\label{def:match}
	Given a slice $p$, 
	the best match to $p$ within the input dataset $I$ is defined as $t = argmax_{q \in I}(cos({\bf p}, {\bf q}))$, where ${\bf p}$ and ${\bf q}$ are the embeddings of slices $p$ and $q$, respectively.
\end{definition}
 
 Practically, an autocompletion tool offers several options for the user to choose from, therefore we are normally interested to look at the top $k$ matches rather than at the best match only. In such a scenario we choose $k$ matches with the highest similarity score to slices ending in $v$.

Once the top $k$ matches are identified, our solution produces a list of recommendations for the next element. It does so by looking at the element that followed each one of the matches in the input dataset. Examples of slices and next elements extracted from the process in Figure~\ref{fig:process}   are shown in Table~\ref{tbl:example_slicing}. In some cases, we may have more than one element following the slice, as we show in the second row of the table.

\setlength{\tabcolsep}{3pt}
\begin{table}
	\centering
	\scriptsize
		\renewcommand\arraystretch{1.3}
		\caption{Sample slices of length $n=3$ for the process in Figure~\ref{fig:process}.}
	\begin{tabular}{|l|}
		\hline
            Slice: Start Event. Task: Check documents. Task: Evaluate. 
			\tabularnewline
			 Next: Exclusive Gateway
			 \tabularnewline
		\hline
            Slice: Task: Check documents. Task: Evaluate. Exclusive Gateway. 
            \tabularnewline
            Next: $[$ Task: Invite to an aptitude test.;  Task: Keep in the applicant pool. $]$
            \tabularnewline
		\hline
			Slice: Task: Evaluate. Exclusive Gateway. Task: Invite to an aptitude test.
			\tabularnewline 
			Next: Exclusive Gateway
			\tabularnewline
		\hline
 			Slice: Exclusive Gateway. Task: Invite to an aptitude test. Exclusive Gateway. 
 				\tabularnewline
 			Next: Task: Rank students according to GPA and the test results
 			\tabularnewline
		\hline
	\end{tabular}
	
	\label{tbl:example_slicing}
\end{table}
\setlength{\tabcolsep}{1pt}

\ignore{
\setlength{\tabcolsep}{3pt}
\begin{table}
	\centering
	\scriptsize
		\renewcommand\arraystretch{1.3}
		\caption{Example of slices of length $n=3$ for the process in Figure~\ref{fig:process}.}
	\begin{tabular}{|l|}
		\hline
            1. Start Event. Task: Check documents. Task: Evaluate.
			\tabularnewline
            2. Task: Check documents. Task: Evaluate. Exclusive Gateway.			
            \tabularnewline
			3. Task: Evaluate. Exclusive Gateway. Task: Invite to an aptitude test.
			\tabularnewline
 			4. Exclusive Gateway. Task: Invite to an aptitude test. Exclusive Gateway.
 			\tabularnewline
			5. Task: Invite to an aptitude test. Exclusive Gateway. Task: Rank students according to GPA\dots
 			\tabularnewline
		\hline
	\end{tabular}
	
	\label{tbl:example_slicing}
\end{table}
\setlength{\tabcolsep}{1pt}
}

Note that the same match $t$ may occur within multiple processes. In such case, the modeler will receive recommendations based on all of the relevant processes. 
As an example, consider the following scenario: processes $A$ and $B$ in the input dataset contain the slice $x \rightarrow y \rightarrow z$. However, in $A$, this slice is connected to node $a$ (that is, $x \rightarrow y \rightarrow z \rightarrow a \subseteq A$), while in $B$, it is connected to $b$ (that is, $x \rightarrow y \rightarrow z \rightarrow b \subseteq B$). In such a case, if a new process $C$ is being constructed, with $x \rightarrow y \rightarrow z \subseteq C$, then if the user asks for a recommendation on what should follow $z$, both $a$ and $b$ will be recommended. Formally, we define a recommendation as follows:

\begin{definition}[Recommendation]
	\label{def:recommendation}
	For a match $t \in I$, we say that the element $r$ is a recommendation, if there is an edge from the last node of $t$ to $r$ in the corresponding process.
\end{definition}
\section{Approach}
\label{sec:approach}

\ignore{
\begin{itemize}
    \item Alternatives for ordering the ``sentences'' which are the task labels
    \item Alternatives for taking gateways/other information into account for prediction. Pros/cons of using them
    \item A running example to show the predictions (preferably from an open dataset)
\end{itemize}
}

Our solution recommends which elements should be added to a process model that is under construction based on an input dataset that is built from a repository of processes. This dataset contains the embeddings for all the slices of predefined length $n$ extracted from those processes. 
The construction of the input dataset is done as follows: for each process, we traverse the graph representing the process in a depth-first order, starting from node $s$, to extract all the slices of length $n$ (see Definition~\ref{def:slice}).
We then compute the embedding vectors for the slices (see Definition~\ref{def:sentence_embedding}).
For each computed embedding we store additional information comprised of the slice itself, a reference to the process from which this slice was extracted, and the elements that followed the slice in the corresponding process, as in Table~\ref{tbl:example_slicing}. 
\ignore{
\begin{algorithm*}[]
	\begin{algorithmic}[1]
		\Function{EmbedDataset}{$I, n$} 
		\Comment{$I$ - dataset of processes, $n \in \mathbb{N}$ - length}
			\State $D \gets \emptyset$  \Comment {Embeddings for $I$}
		\For {$G=\{V, E, s, t\} \in I$}  
		\State $S \gets $  all slices of length $n$ in G
		\Comment{(Definition~\ref{def:slice})}
	
		\For {$p \in S$}  
		    \State $\textbf{p} = U(p)$ \Comment {Compute embedding (Definition~\ref{def:sentence_embedding})}
		    \State $D \gets D   \cup \textbf{p}$ 
		\EndFor
		\EndFor
	
		\State\Return $D$ 
		\EndFunction     

	\end{algorithmic}
	\normalsize
	\caption{Pseudo-code for computing embeddings for the input dataset.}
	\label{alg:dataset_embedding}
\end{algorithm*}
\vspace{-0.2cm}
}

During the autocompletion phase, our solution follows the steps presented in Algorithm~\ref{alg:recommendation}. It receives as one of its parameters the 
node $v$ for which it needs to recommend the next element(s). It first determines what are the slices that lead to node $v$ (that is, end in node $v$). This is done in routine \textit{ExtractSlices} (line~\ref{row:L}), which traverses the graph starting from node $v$ backwards, based on the incoming edges. 
Note that we only attempt to make a recommendation if we have at least one such slice. 

Next, our recommender computes the embedding for each extracted slice. All the embeddings are stored in matrix $X$ (line~\ref{row:X}). We then compute the similarity matrix between $X$ and the input set embeddings stored in $D$ (line~\ref{row:M}). 

Our algorithm proceeds in line~\ref{row:T}, where we find the top $k$ matches for the slices from $S$ with \textit{ExtractTopMatches} routine. This routine looks for the best match based on the highest similarity score, records the matching slice and corresponding next-element recommendations, and repeats until $k$ recommendations are collected and presented to the end user. In case that we have the same recommendation for two different matches, we present it only once. 

\begin{algorithm*}[h]
	\begin{algorithmic}[1]
		\Function{RecommendElement}{$G, v, n, k, D$} 
		\Comment{$G$ - process (graph), $v$ - node in $G$ (element), $n \in \mathbb{N}$ - length, $k$ - maximum number of top recommendations, $D$ - input dataset embeddings }
		\State $S \gets ExtractSlices(G, v, n)$ 
	\label{row:L}
	
		\Comment{Extract slices of length $n$ that terminate in $v$ (Definition~\ref{def:slice})}
		\State $X \gets \emptyset$  \Comment {Embeddings for slices in $S$}
		\State $R \gets \emptyset$  \Comment {Recommended elements}

		\For {$p \in S$}  
		    \State $\textbf{p} \gets U(p)$ \Comment {Compute embedding (Definition~\ref{def:sentence_embedding})}
		    \State $X \gets X   \cup \textbf{p}$ \label{row:X}
		\EndFor
		
		\State $M \gets M(X, D)$ \Comment{Compute similarity matrix (Definition~\ref{def:sim_matrix})} \label{row:M}
		\State $T \gets ExtractTopMatches(M, k)$ \Comment{Extract $k$ most similar entries in $D$ to $S$} \label{row:T}
		\For {$t \in T$} \label{row:rstart} 
		    \State $r \gets$ Get recommendations for $t$ \Comment{(Definition~\ref{def:recommendation})}
		    \State $R \gets R   \cup r$   
		\EndFor
		\State\Return $R$  \label{row:rend}
		\EndFunction

	\end{algorithmic}
	\normalsize
	\caption{Pseudo-code for the autocompletion step. Given process $G$ and element $v$, recommend what other element(s) should be added to $G$ after $v$.}
	\label{alg:recommendation}
\end{algorithm*}

\section{Evaluation}
\label{sec:eval}

\subsection{Datasets}

We evaluated our approach on four datasets of process models from different domains. These are summarized in Table~\ref{tbl:datasets}.
The first dataset (labeled \airport{}) is comprised of $24$ processes capturing airport procedures~\cite{apromore}.
The second dataset contains $40$ processes available from the Integrated Adaptive Cyber Defense (\iacd{}) initiative~\cite{iacd}. These processes are used for modeling  security automation and orchestration workflows. 
In \iacd{}, almost $85\%$ of the tasks are unique, that is, they appear only once in the entire dataset.
The third dataset (labeled \proprietary{}) is a collection of proprietary processes used in a product for security orchestration. It resembles the \iacd{} dataset and contains $18$ models. Unlike \iacd{}, the processes in the proprietary dataset reuse almost one third of the elements among them.
The last dataset (\university{}) is used to model university admission procedures for master students in German universities~\cite{antunes2015process}. The dataset is comprised of $9$ models, with $40\%$ of the same labeled elements occurring in at least two models.

\setlength{\tabcolsep}{1pt}
\begin{table}
	\centering
	\scriptsize
		\renewcommand\arraystretch{1.3}
		\caption{Characterization of the datasets used in the experiments.
		For each dataset, from left to right, we show its number of processes, 
		the number of elements in each process, how many of the elements are unique and appear only once in the dataset,
		the number of slices extracted for a length of~\slice{},
		the average number of elements per process,
		the percentage of elements that are used in more than one process,
		and the average number of slices per element.}
	\begin{tabular}{|l|c|c|c|c|c|c|c|}
		\hline
	         Dataset &  Processes  & Elements & Unique & Slices  & Elements/Process & Shared Elements & Slices/Element
			  \tabularnewline
		\hline\hline

\airportt{}  & 24 & 984  & 655 & 37868 & 41 & 33.5\% & 38.5
			\tabularnewline
		\hline
		
\iacdt{}  & 40 & 676  & 571 & 1451 & 16.9 & 15.5\% & 2.1 
			\tabularnewline
		\hline
\proprietaryt{}  & 18  & 508 & 348 & 1841 & 28.2& 31.5\%  & 3.6 
			\tabularnewline
		\hline
			\universityt{} & 9  & 451 & 272 & 10050 & 50.1 & 40\% &  22.3 
			\tabularnewline
		\hline
	\end{tabular}
	
	\label{tbl:datasets}
\end{table}
\setlength{\tabcolsep}{1pt}

The efficiency of a recommendation system depends heavily on the data distributions of the input dataset and the process that is being developed. Simply put, if the processes are completely unrelated to each other, then it is very difficult to make meaningful suggestions. On the other hand, if the processes share semantic meaning, then our recommendation engine can leverage that and provide useful suggestions to the end-user.

The datasets used in the experiment have quite different characteristics that affect the ability of our autocompletion engine to make good recommendations. 
For example, in \iacd{} only $15.5\%$ of the elements are shared among the processes. Also, the processes in \iacd{} are relatively small, as captured by the ``Elements/Process'' column, which means that during the input dataset construction, we can extract only a small number of slices.

In contrast, all of the processes in the \university{} dataset cover similar procedures, thus we expect a significant amount of reuse. Even when there are elements that have somewhat different labels (and are therefore counted as ``Unique'' elements), they often bear similar meanings, e.g., ``Accept'' task label is similar to ``Send letter of acceptance''. The processes here and in the \airport{} dataset are also larger than in the other two datasets, allowing our auto-completion engine to mine a more meaningful and diverse input dataset to be used for its recommendations.

\vspace{-0.4cm}
\subsubsection{Validation methodology.}
We use the leave-one-group-out cross-validation technique to evaluate the accuracy of our approach. 
Each process, in turn, is evaluated against the rest of the processes that serve as the input dataset to mine recommendations from. This gives us an opportunity to use all available process models as an input and as a validation dataset.
For each cross-validation fold, we choose a different process model and generate sub-processes with an increasing number of elements, in order to simulate different stages of the model construction. 
For each sub-process, we also record the next element's type and label that should be recommended (the ground truth).
At each evaluation step, we perform the autocompletion step of Algorithm~\ref{alg:recommendation} for each one of the sub-processes and obtain different accuracy metrics comparing the top recommendations against the ground truth.
In Section \ref{subsec:experiments} we present the averages of the accuracy metrics that we obtain from the cross-validation folds.

\subsection{Metrics}
\label{subsec:metrics}

In our evaluation we use two types of metrics; one type focuses on evaluating the precision and recall of the recommendations, and a second type assesses the quality of the predictions that are semantically similar to the ground truth.


For the first type, we use \texttt{precision@k} and \texttt{recall@k}~\cite{hawking2001measuring}. \texttt{Precision@k} is the fraction of elements in the top $k$ recommendations that match the ground truth, while \texttt{recall@k} is the coverage of the ground truth in the top $k$ recommendations. Note that this type of metrics only captures cases where the labels of the elements recommended by our solution match the ground truth precisely. In our experiments we used $k=3$.

For the second type, we focus on the metrics \bleu{}~\cite{papineni2002bleu} and \meteor{}~\cite{lavie2007meteor}, that are frequently used to evaluate machine translation systems~\cite{mathur2020results}. 
\bleu{} is the precision of $n$-grams of a machine translation's output compared to the ground truth reference. This metric is weighted by a brevity penalty to compensate for recall in overly short translations. 
\meteor{} is similar to \bleu{} but takes explicit ordering of words into account. During its matching computation it also considers translation variability via word inflection variations, synonym and paraphrasing.
Additionally, we report the \cosine{} similarity of the predictions to the ground truth, computed based on Definition~\ref{def:cosine}. This type of metrics captures both exact and similar matches between recommendations and the ground truth.

Table~\ref{tbl:metrics} shows values of \cosine{} similarity, \bleu{} and \meteor{} for some sample recommendations to gain intuition on these metrics. 
Note that higher scores of \bleu{}, \meteor{}, and  \cosine{} are correlated with higher semantic similarity of sentences. 
\bleu{} and \meteor{}'s range is $[0, 1]$, while \cosine{} similarity spans from $-1$ to $1$. Values over $0.3$ represent understandable to good translations~\cite{bleuexplained} for both \bleu{} and \meteor{}; values over $0.4$ represent high quality translations, and they exceed $0.5$ for very-high quality translations. Values of $0.2$ to $0.3$ represent cases where one could see the gist of the translation, but it is not very clear.


\setlength{\tabcolsep}{2pt}
\begin{table}
	\centering
	\scriptsize
		\renewcommand\arraystretch{1.3}
			\caption{\cosine{} similarity, \bleu{} and \meteor{}  values for sample recommendations and ground truth values from the \university{} 
			dataset~\cite{antunes2015process}.}
	\begin{tabular}{|c|c|c|c|c|c|}
		\hline
	        Recommendation & Ground Truth & \cosinet{} & \bleut{} & \meteort{} 
			  \tabularnewline
		\hline\hline
		Send letter of acceptance & Send letter of provisional acceptance & 0.82 & 0.80 & 0.91
			\tabularnewline
		\hline
		Attach additional requirements & Collect additional required documents & 0.58 & 0.35 & 0.60
 			\tabularnewline
 		\hline
Send interview invitation & Invite for talk & 0.60 & 0 & 0.17
			\tabularnewline
		\hline		
	
Check bachelor's degree & Wait for bachelor's certificate & 0.62 & 0.25 & 0.16 
 			\tabularnewline
 		\hline
	\end{tabular}

	\label{tbl:metrics}
\end{table}
\setlength{\tabcolsep}{1pt}


\subsection{Experiments}
\label{subsec:experiments}

We first study the effect that the selected slice length has on the quality of the recommendations made by our autocompletion engine. We then compare our solution to a random algorithm that makes recommendations based on the statistical distribution of the elements in the input dataset.

At the beginning, we allowed the algorithms to make predictions based on all available elements in the processes. We learned quickly that the majority of the predictions (over $80\%$) were for ground truth of end events and gateways. Clearly, one may suggest an autocompletion engine that always suggests to use these two types of elements and get quite good precision and recall. However, this is not very useful for the model designer. We therefore also checked the quality of the suggestions when gateways and end events were excluded from recommendations and the ground truth (we refer to this case as \texttt{Filtered}).

\vspace{-0.4cm}
\subsubsection{Slice length study.} 
\begin{figure}[htb]
    \centering
\begin{subfigure}{0.48\textwidth}
  \includegraphics[width=\linewidth]{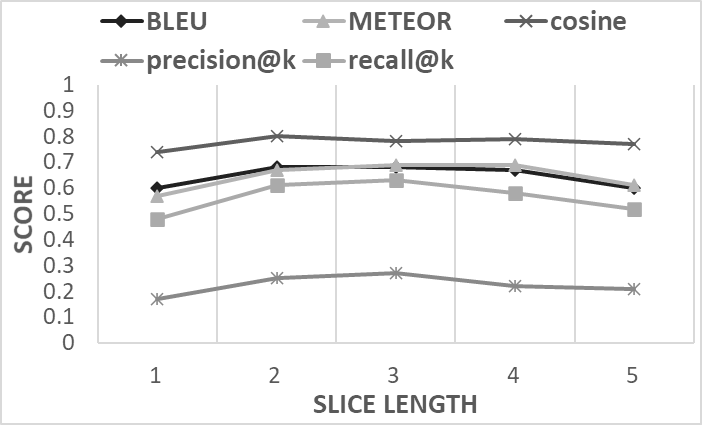}
  \caption{\airportt{} dataset}
  \label{fig:window_study_airport}
\end{subfigure}
\begin{subfigure}{0.48\textwidth}
  \includegraphics[width=\linewidth]{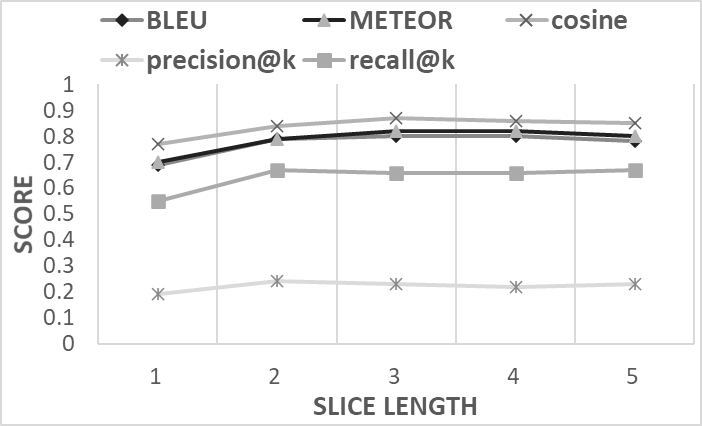}
  \caption{\iacdt{} dataset}
  \label{fig:window_study_iacd}
\end{subfigure}\hfil 
\vspace{8pt}
\medskip
\begin{subfigure}{0.48\textwidth}
  \includegraphics[width=\linewidth]{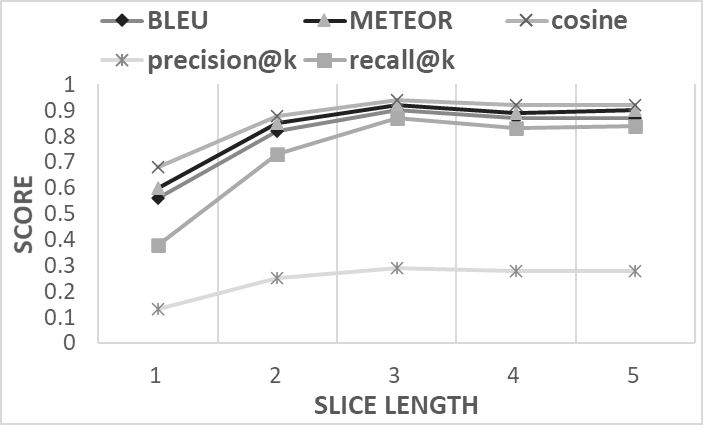}
  \caption{\proprietaryt{} dataset}
  \label{fig:window_study_proprietary}
\end{subfigure}
\begin{subfigure}{0.48\textwidth}
  \includegraphics[width=\linewidth]{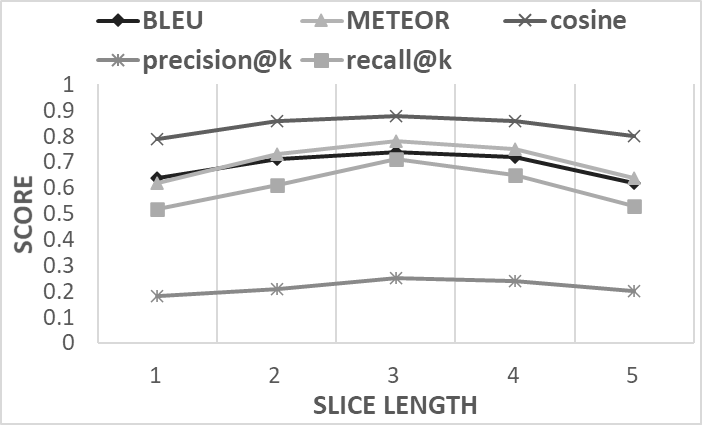}
  \caption{\universityt{} dataset}
  \label{fig:window_study_university}
\end{subfigure}

\caption{Evaluation metrics for slice lengths of $1$ to $5$, $k=3$, for \texttt{Filtered} case.}
\label{fig:window_study}
\end{figure}

We varied the length of the slice from $n=1$ to $n=5$ and collected the metrics for each dataset. 
Figure~\ref{fig:window_study} shows the results of this experiment for the \texttt{Filtered} case. The graphs in the figure plot 
the slice lengths on the x-axis and metric scores on the 
y-axis for the five metrics discussed in Section~\ref{subsec:metrics}.
On average\footnote{Computed as average over all four datasets, but not shown due to space limitation.}, the best results are for a slice length of $n=3$, although the effect of the slice length on each dataset varies. 

For \airport{} (Figure \ref{fig:window_study_airport}) and \university{}  (Figure \ref{fig:window_study_university}) datasets, our autocompletion engine's accuracy increases and then decreases slightly as the slice gets longer, with the best results observed at $n=3$. 
The \iacd{} dataset (Figure \ref{fig:window_study_iacd}) is the most stable of all the four datasets with respect to the effect of the slice length on the algorithm.
For the \proprietary{} dataset (Figure \ref{fig:window_study_proprietary}) the accuracy improves till $n=3$ and then remains stable.

For the case where all the elements are taken into consideration\footnote{Not visualized due to space limitation.}, slice length of $n=2$ gives slightly better results than $n=3$ for some datasets. We use slice of length $n=3$ throughout the rest of the experiments to enable easy comparison. However, we observe that one may need to carry out a preliminary slice length study for each new domain/dataset to get the best recommendations.
This can be done automatically during the input dataset construction process.

\vspace{-0.4cm}
\subsubsection{Comparison to a random algorithm.}

In this experiment we studied the efficiency of our solution (labeled \slicing{}) in comparison to an algorithm that autocompletes the process at random (labeled \random{}). We collected the occurrences of each element in the dataset, and randomly selected the top recommendations based on the statistical distribution of the elements. Elements that occurred more frequently had a higher chance of being selected.

We set the slice length for our algorithm to $n=3$ based on the previous experiment. The random algorithm's performance is agnostic to this choice, as it takes no slices into account. We executed the experiment with \random{} $30$ times, and computed averages over all runs.





\definecolor{LightCyan}{rgb}{0.88,1,1}

\setlength{\tabcolsep}{1pt}
\begin{table}
    \centering
    \renewcommand\arraystretch{1.3}
     \scriptsize
    \caption{Evaluation metrics computed for \slicing{} and \random{}, for all the elements (top half) and for filtered out gateways and end events (bottom half).
    }
    \begin{tabular}{|c|c|c|c|c|c|c|c|c|}
		\hline
		\rowcolor{LightCyan}	\multicolumn{9}{|c|}{ All elements} \\
		\hline
	    Dataset/ & 	      \multicolumn{2}{c|}{\airportt{}}  & \multicolumn{2}{c|}{\iacdt{}} & \multicolumn{2}{c|}{\proprietaryt{}} &
	      	      \multicolumn{2}{c|}{\universityt{}} 
	      \\ 
	     \cline{2-9}
	     Metric & \slicingt{} & \randomt{}  & \slicingt{} & \randomt{} & \slicingt{} & \randomt{} &  \slicingt{} & \randomt{} \\
		\hline
		\hline
				\bleut{}& 0.56$\pm$0.27 & 0.37$\pm$0.29    & 0.71$\pm$0.33 & 0.29$\pm$0.22 & 0.7$\pm$0.29 & 0.37$\pm$0.28 & 
				0.64$\pm$0.24 & 0.44$\pm$0.28  \\
		\hline
		 \meteort{}  & 0.63$\pm$0.4 & 0.34$\pm$0.38    & 0.75$\pm$0.34 & 0.26$\pm$0.26 & 0.81$\pm$0.33 & 0.4$\pm$0.37 & 
		 0.77$\pm$0.34 & 0.44$\pm$0.39\\
		 	\hline
		 \cosinet{}  & 0.8$\pm$0.31 & 0.6$\pm$0.34 & 0.82$\pm$0.27 & 0.43$\pm$0.23 & 0.89$\pm$0.24 &  0.63$\pm$0.31  & 
		 0.87$\pm$0.26 & 0.64$\pm$0.32    \\
		 	\hline
		 			 \precisiont{} & 0.21$\pm$0.18 &  0.09$\pm$0.15 & 0.22$\pm$0.17 & 0.02$\pm$0.08 & 0.28$\pm$0.14 & 0.1$\pm$0.15 & 
		 			 0.29$\pm$0.18 & 0.13$\pm$0.16   \\
		 			 	 	\hline
		 			 \recallt{} & 0.61$\pm$0.49 &  0.27$\pm$0.44 & 0.63$\pm$0.48 & 0.06$\pm$0.25 & 0.81$\pm$0.33 & 0.3$\pm$0.46 &
		 			 0.78$\pm$0.34 & 0.4$\pm$0.49   \\
		 	\hline
		 	\hline
		\rowcolor{LightCyan}	\multicolumn{9}{|c|}{Filtered} \\
		\hline
	  	Dataset/ &
	     	      \multicolumn{2}{c|}{\airportt{}} 
	      & \multicolumn{2}{c|}{\iacdt{}} & \multicolumn{2}{c|}{\proprietaryt{}} &
	      	      \multicolumn{2}{c|}{\universityt{}}
	      \\ 
	     \cline{2-9}
	     Metric & \slicingt{} & \randomt{}  & \slicingt{} & \randomt{} & \slicingt{} & \randomt{} & \slicingt{} &  \randomt{} \\
		\hline
		\hline
			\bleut{} & 0.68$\pm$0.32  & 0.3$\pm$0.13 & 0.8$\pm$0.28 & 0.32$\pm$0.18 & 0.9$\pm$0.24 & 0.21$\pm$0.2 & 
			0.74$\pm$0.28 & 0.35$\pm$0.21   \\
		\hline
		 \meteort{}  & 0.69$\pm$0.38 &  0.15$\pm$0.14 & 0.82$\pm$0.27 & 0.29$\pm$0.22 & 0.92$\pm$0.21 & 0.21$\pm$0.23& 
		 0.78$\pm$0.34 & 0.2$\pm$0.21  \\
		 	\hline
		 \cosinet{} & 0.78$\pm$0.27 & 0.4$\pm$0.14 & 0.87$\pm$0.21 & 0.44$\pm$0.18 & 0.94$\pm$0.16 & 0.39$\pm$0.2 & 
		 0.88$\pm$0.21 & 0.47$\pm$0.18  \\
		 	\hline
		 			 \precisiont{} & 0.27$\pm$0.27 &  0$\pm$0.04 & 0.23$\pm$0.17 & 0.01$\pm$0.06 & 0.29$\pm$0.12 & 0.01$\pm$0.06 & 
		 			 0.25$\pm$0.18 & 0.01$\pm$0.06 \\
		 			 	 	\hline
		 			 \recallt{} & 0.63$\pm$0.48 &  0.01$\pm$0.11 & 0.66$\pm$0.47 & 0.03$\pm$0.17 & 0.87$\pm$0.34 & 0.03$\pm$0.17 & 
		 			 0.71$\pm$0.45 & 0.03$\pm$0.17   \\
	
		\hline
	\end{tabular}
    
\label{tbl:eval}
\end{table}

\begin{figure}[h]
    \centering
\begin{subfigure}{0.48\textwidth}
  \includegraphics[width=\linewidth]{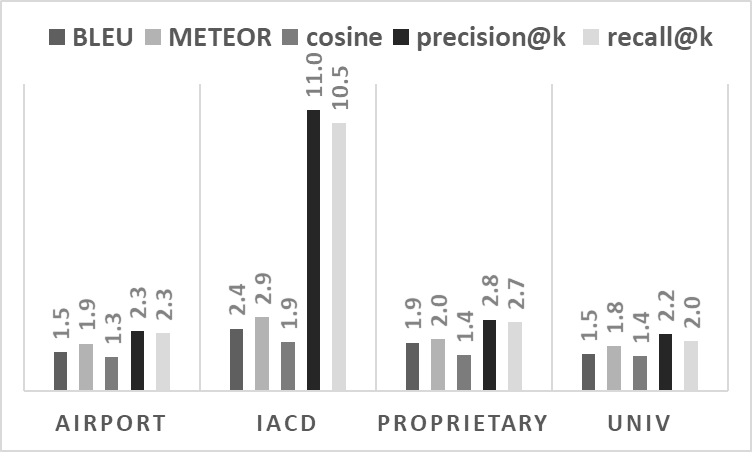}
  \caption{Slicing/Random ratio, all elements}
  \label{fig:slicing_vs_random_all_nodes}
\end{subfigure}\hfil 
\begin{subfigure}{0.48\textwidth}
  \includegraphics[width=\linewidth]{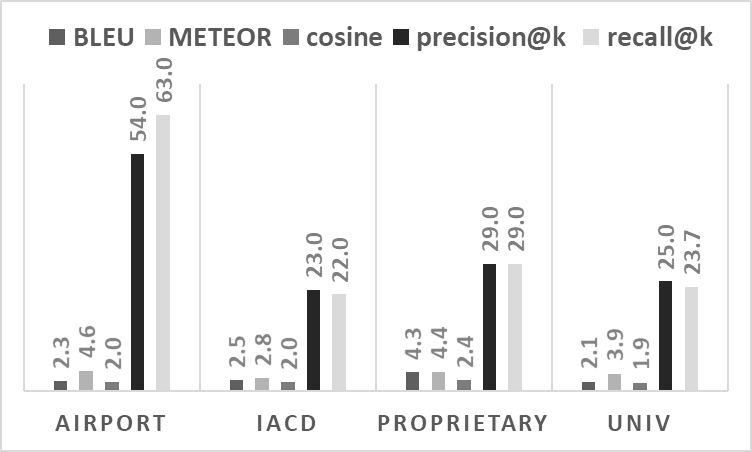}
  \caption{Slicing/Random ratio, filtered}
  \label{fig:slicing_vs_random_filtered}
\end{subfigure}
\caption{Ratio between the values of the metrics measured for the \slicing{} versus the \random{} algorithm. \texttt{UNIV} stands for the \university{} dataset.}

\label{fig:eval_ratio}
\end{figure}

The results of this experiment are shown in Table~\ref{tbl:eval} and Figure~\ref{fig:eval_ratio}. Table~\ref{tbl:eval} gives the values of the metrics computed for our algorithm and for the random algorithm, for the two different configurations (including all the nodes, or filtering out gateways and end events). We use the $\pm$ notation to present the average and the standard deviation for each metric. 
In Figure~\ref{fig:eval_ratio} we visualize the ratio between the averages of the metrics computed for the two algorithms for easier comparison.\footnote{Some precision and recall values are rounded to $0$ when only two decimal places are used. For such cases, we use higher precision values to compute the ratio.}

When all the elements are taken into account, our algorithm achieves scores over $0.56$ for  \bleu{} and \meteor{}  metrics, which indicates a very good match between the ground truth and the recommendations~\cite{bleuexplained}. \cosine{} similarity is also high, especially for the \university{} dataset.
Our algorithm  performs much better than the random algorithm with respect to \bleu{}, \meteor{} and \cosine{} similarity. It also has a much higher precision and recall. Note that since at every step we recommend the top \topk{} elements, then a precision of $0.29$ (for the \university{} dataset, with all the elements, and for the \proprietary{ dataset, in the \texttt{Filtered}} case) means that, in average, we suggest one element out of \topk{} correctly almost always. The random algorithm reaches up to $0.13$ precision when all the elements are taken into account. There are many gateways and end elements in both the input and the validation datasets, so the random algorithm suggests them frequently, and, therefore, gets many accurate predictions.   

When we narrow down the analysis only to activity tasks, we learn that our algorithm outperforms the random algorithm for all the datasets, as witnessed by all the metrics. Its precision is up to $64$x higher than that of the random algorithm, with the recall having up to $63$x improvement.  

\subsection{Findings}
We can distill the following key findings from our empirical evaluation:
\begin{itemize}[leftmargin=*]
\item \emph{Our autocompletion engine is applicable to various domains}: The recommendations obtained for the presented datasets are quite accurate as all the metrics attest to. Precision of $0.21$ to $0.29$ means that in average one out of three to five suggestions is correct. 
Since each recommendation contains three options, it means that most recommendations include one exact match to the ground truth. A recall of up to $0.87\%$ indicates that we manage to cover the majority of the expected elements in our recommendations. Even if a suggestion is not an exact match to the ground truth, the high values of the \bleu{}, \meteor{}, and \cosine{} metrics indicate a significant similarity between them. 
    
    \item \emph{Slicing window length has a mild impact on the quality of autocompletion}: The results from our first experiment show that our recommendation engine can indeed get better predictions for some slice sizes. We therefore recommend tuning this parameter during the input dataset construction procedure for each new dataset. However, metrics measurement changes are not drastic enough for us to claim that this parameter has a significant impact on our algorithm.
    \item \emph{Semantic similarity based evaluation metrics exhibit mostly high correlation among them}: The results from both the slice length study and the comparison of \slicing{} to \random{} 
    show that the \meteor{} and \bleu{} metrics have a Pearson correlation of over $0.85$, except when the \random{} algorithm is applied in the \texttt{Filtered} case. We witnessed similar results for \meteor{} and \cosine{}, with a Pearson correlation of over $0.88$. The semantic similarity based metrics had a low correlation to the metrics used to evaluate exact matches.
    
    The main difficulty in selecting the right metrics is related to the fact that none of them are tuned specifically to the problem at hand. For example, measuring precision based on exact matches does not reveal the full potential of our algorithm, as predictions similar to the ground truth (e.g., ``Accept'' versus ``Send letter of acceptance'') are not considered as a match. 
    A fruitful area of future research would be to study to what extent these metrics correlate with the feedback of subject matter experts on the quality of the recommendations.
\end{itemize}

\subsection{Threats to validity}

One threat to validity is related to the choice of \use{} for the embeddings computation. The version we used has been trained on English sentences. This means that we can not use it for models in other languages. In the future, we plan to investigate whether multi-language models or other encoders may improve our autocompletion solution.

Another threat is related to the difficulty in comparing our results to those made by other researchers, due to lack of evaluation on open source datasets. We hope our work can serve as a baseline for such comparisons.

\section{Related Work}
\label{sec:related}

Similarity between process models has been a subject of interest for a while now. The main goal of this line of research is to determine if two processes are similar to each other, or implement a query to find a process, rather than recommending how to autocomplete a process.

One of the first attempts to combine structural and semantical information while assessing models similarity was 
accomplished with BPMN-Q~\cite{awad2008semantic}.  BPMN-Q allows  users  to  formulate structure-related  process  model  queries  and uses WordNet~\cite{wordnet} knowledge to make the search semantic aware. WordNet-based word-to-word semantic similarity values were also used in~\cite{shahzad2018wordnet} for this task. 

Later, the concept of a causal footprint was introduced~\cite{van2013measuring}. 
A causal footprint is a collection of behavioral constraints imposed by a process model. These constraints are converted into vectors and are compared via a similarity score. 


Some researchers focused on detecting similarities between processes based on graph edit distances between them~\cite{ivanov2015bpmndiffviz,dijkman2009graph}, while others~\cite{starlinger2014similarity} used a combination of graphs' structure related techniques with per-element similarity.  


Researchers leveraged various process similarity techniques to enable process autocompletion. One such example is the tool  FlowRecommender~\cite{zhang09}, that suggests next elements based on pattern matching for graphs. It was evaluated on a synthetic dataset and only focused on validating the structural accuracy of the prediction. As an improvement to this technique, the authors proposed to traverse the process graphs, and compare them via a string edit distance similarity metric~\cite{li14}. The approach focused on finding isomorphic graphs when labels had to match precisely.

In a more recent work~\cite{wang2019process}, the authors used bag-of-words to predict the top k most similar process models from a process repository to a process that is being modeled. 
They also used a structural-only based comparison in~\cite{wang2019ks} to detect differences between models. Unfortunately, the authors did not evaluate their approach on any open dataset using any known metrics, making it difficult to compare their approach to ours.


Machine learning based techniques for processes included graph embeddings to search processes with a search query that contained arbitrary text~\cite{Yu2020}. This work focused on a single node matching against the search query. Earlier, transfer learning enabled search and retrieval of processes from logs~\cite{koohi2018cross}, where unique content-bearing workflow motifs were extracted from the set of processes. These motifs were treated as features and then each process was represented as a vector in this feature space. 
Based on this representation, similarity metrics can be computed between processes, and used in the future to improve our solution.

In an alternative method for encoding process models as vectors~\cite{de2018act2vec},
De Koninck et. al. developed representation learning architectures for embedding trace logs and models to enable comparison between them. The learned architecture focused on the structural properties of processes, 
rather than their semantics. 

Finally, Burgue\~no et. al.~\cite{burgueno:hal-03010872} proposed to use contextual information taken from process description to auto-generate the process. Indeed such documentation, when available, could be used to improve our autocompletion technique.
\section{Conclusions and Future Work}
\label{sec:conclusions}

\ignore{
\begin{itemize}
    \item To the best of our knowledge, this work is the first attempt to recommend business processes based on USE 
    \item Our technique for building evaluation test sets can be reused while evaluating other types of recommendation systems for business processes
    \item Our evaluation shows that the predictions are fast and accurate (provide supporting arguments from the evaluation section)
\end{itemize}
}

In this paper, we presented a novel technique for design-time next-task recommendation in business process modeling that is based on semantic similarity of processes. Our solution supports business modelers
by autocompleting the next element(s) during process construction. We used a state-of-the-art NLP technique that detects similarities between sentences and adopted it to the domain of business processes models. This allowed us to overcome the challenge of having little data to train a traditional machine learning recommender model on.

Our evaluation shows that the suggestions made by our recommendation engine are accurate for datasets with different characteristics and from different domains. Moreover, our solution is suitable for applications in commercial products, as the evaluation on a proprietary dataset shows.

In the future, we plan to conduct a user study where process modelers will rate the predictions made by our tool. This will allow us to better assess the efficiency of our approach in practical settings and also learn which metrics are most suitable for the evaluation of other process recommendation systems.

Another interesting direction would be to investigate how the BPMN execution semantic could be taken into account. By analyzing execution traces of processes from the input dataset and  encoding the traces in addition to paths.

\bibliographystyle{splncs04}
\bibliography{refs}
\end{document}